\documentclass[12pt]{iopart}
\usepackage[english]{babel}
\usepackage[utf8]{inputenc}
\usepackage{iopams}
\usepackage{bbm}
\usepackage{graphicx}
\usepackage{braket}
\usepackage[colorlinks, allcolors=blue]{hyperref}
\usepackage{cite}

\begin{document}

\newcommand{\norm}[1]{\left\lVert#1\right\rVert}

\newcommand{\mean}[1]{\ensuremath{\left\langle #1 \right\rangle}}
\newcommand{\matrixel}[3]{\ensuremath{\left\langle #1 \middle| #2 \middle| #3 \right\rangle}}
\newcommand{\de}{\mathrm{d}}
\newcommand{\Ham}{\mathcal H}

\title[Loschmidt echo singularities as dynamical signatures of strongly localized phases]{Loschmidt echo singularities as dynamical signatures of strongly localized phases}

\author{Leonardo Benini$^{1,2}$, Piero Naldesi$^{3,4}$, Rudolf A. Römer$^{1,5}$, Tommaso Roscilde$^2$}
\address{$^1$
Department of Physics, University of Warwick, Coventry, CV4 7AL, UK%
}
\address{$^2$
Univ de Lyon, Ens de Lyon, Univ Claude Bernard,
 and CNRS, Laboratoire de Physique, F-69342 Lyon, France}
\address{$^3$
Universit\'e  Grenoble-Alpes,  LPMMC and  CNRS, F-38000  Grenoble,  France}
\address{$^4$
Institute for Quantum Optics and Quantum Information of the Austrian Academy of Sciences, Innsbruck, Austria}
\address{$^5$
CY  Advanced  Studies  and  LPTM  (UMR8089  of  CNRS),CY  Cergy-Paris  Universit\'e,  F-95302  Cergy-Pontoise,  France}

\begin{abstract}

Quantum localization (single-body or many-body) comes with the emergence of local conserved quantities --- whose conservation is precisely at the heart of the absence of transport through the system.
In the case of fermionic systems and $S=1/2$ spin models, such conserved quantities take the form of effective two-level systems, called $l$-bits. While their existence is the defining feature of localized phases, their direct experimental observation remains elusive. Here we show that strongly localized $l$-bits bear a dramatic universal signature, accessible to state-of-the-art quantum simulators, in the form of periodic cusp singularities in the Loschmidt echo following a quantum quench from a N\'eel/charge-density-wave state. Such singularities are perfectly captured by a simple model of Rabi oscillations of an ensemble of independent two-level systems, which also reproduces the short-time behavior of the entanglement entropy and the imbalance dynamics. In the case of interacting localized phases, the dynamics at longer times shows a sharp crossover to a faster decay of the Loschmidt echo singularities, offering an experimentally accessible signature of the interactions between $l$-bits.

\end{abstract}


\submitto{\NJP}
\maketitle

\section{Introduction}
Constructive interference of paths bringing a particle back to its initial location in real space is at the heart of single-particle (or Anderson) localization (AL)\cite{Akkermans-book, Anderson-book}; more recently a similar phenomenon occurring in Hilbert space (MBL) \cite{NandkishoreH2015,Alet2018,Abaninetal2019, GOPALAKRISHNAN2020} has been shown to prevent many-body quantum systems from relaxing to thermal equilibrium, undermining the ergodic hypothesis in a large class of models of interacting quantum particles. 
Localized phases are generally characterized in the negative (absence of transport, of long-range order, of spectral gaps, etc.), while positive characterizations are typically elusive. A crucial aspect of localization is the persistence of initial conditions, which, in the case of AL of non-interacting particles, is related to the conservation of populations in the localized single-particle eigenstates of the Hamiltonian. In the case of MBL, the analog phenomenon would be the appearance of local conserved quantities (called local integrals of motion or $l$-bits \cite{Serbynetal2013,Huseetal2014,Imbrie2016, Imbrieetal2017}) which are obtained by unitary transformations of local operators; and which, if extensive in number, constrain the dynamics of the system to the point of preventing relaxation. 

The existence of $l$-bits in disordered spin chains can be mathematically proven under the assumption of limited level attraction \cite{Imbrie2016}, and approximate $l$-bits for many-body systems can be constructed with a variety of analytical as well as numerical methods \cite{Rosetal2015, RademakerO2016,Youetal2016,InglisP2016,Obrienetal2016,Pekkeretal2017,Goihletal2018,Kulshreshthaetal2018,Mierzejewskietal2018,Pengetal2019}. Much of the phenomenology of MBL dynamics (persistence of traits of the initial state, logarithmic growth of entanglement entropies, etc.) -- observed in numerical studies as well as in experiments \cite{Alet2018,Abaninetal2019}  -- can be directly explained in terms of the existence of $l$-bits and interactions between them. Yet observing $l$-bits directly is an arduous task, given that their expression is highly disorder-dependent (and generally unknown even in theory), and it would require high-precision measurements of local observables in different local bases. An even more ambitious task for experiments is the one of probing directly the existence of interactions among $l$-bits, which is a defining feature distinguishing MBL from AL. Measuring the consequences of such interactions on correlation and entanglement dynamics is currently the focus of a considerable experimental effort based on state-of-the-art quantum simulation platforms \cite{Rispolietal2019, Lukin2019, Chiaroetal2020}.

The purpose of this work is to show that, in the case of strongly localized phases, the existence of $l$-bits can offer striking signatures in the dynamics of the Loschmidt echo, namely in the logarithm of the return probability to the initial state $|\psi_0\rangle$
\begin{equation}
\lambda(t) = - \frac{1}{L} \left [ \log |\langle  \psi_0 | e^{-i {\cal H} t} |\psi_0\rangle|^2\right ]_{\rm av}~.
\label{e.LE}
\end{equation}
Here ${\cal H}$ is the system's Hamiltonian and $L$ the lattice size; $[...]_{\rm av}$ indicates the disorder average. Please note that we employ the ``Loschmidt echo'' terminology similar to other recent works in which singular behavior has been observed in the quench dynamics of many-body quantum systems~\cite{Heyl2013, Heyl2014,Jurcevicetal2017, Yang2017, Yin2018, Halimeh2018, Heyl2018, Guoetal2019}. Nonetheless, In the quantum-chaos literature the Loschmidt echo is more generally defined as the scalar product between the evolution of the same state $\ket{\psi_0}$  with two different Hamiltonians, $H_1$ and $H_2$~\cite{Peres1984, Jalabert2001}. Starting from $\mathcal{L}(t)=\braket{\psi|e^{iH_2t}e^{-iH_1t}|\psi}$, our definition is retrieved in the case of $H_1=\mathcal{H}$ and $H_2=0$. When $|\psi_0\rangle$  has a simple factorized form, and in the case of strong disorder, we find that the Loschmidt echo displays periodic singularities, decaying very slowly in amplitude -- as illustrated using a model of disordered spinless fermions in 1d (corresponding to the $S=1/2$ XXZ model in a fully random or quasi-periodic field) initialized in a charge density-wave (CDW) state. The singularities in the Loschmidt echo are fully explained quantitatively by a simple model of a collection of localized 2-level systems (2LS) undergoing independent Rabi oscillations, and approximating strongly localized $l$-bits. The same minimal model captures quantitatively the dynamics of the entanglement entropy at short times as well as of the number entropy at longer times; and the dynamics of the density imbalance characterizing the initial state. 

At longer times the deviation of the exact results for the MBL dynamics from the predictions of the 2LS ensemble offers direct evidence of the interactions among the $l$-bits in the form of a faster decay of the Loschmidt-echo singularities and imbalance oscillations. As the Loschmidt echo and the imbalance are generally accessible to quantum simulators, either measuring individual degrees of freedom \cite{Gaerttneretal2017, Jurcevicetal2017} or even global ones \cite{Schreiberetal2015, Bordia2015, Luschenetal2017}, our results show that strong direct signatures of $l$-bits dynamics and interactions are within the immediate reach of state-of-the-art experiments on disordered quantum systems.

The structure of the paper is as follows: section~\ref{s.Model} introduces the XXZ model in a random/quasi-periodic field; section~\ref{sub:LESING} discusses the observation of Loschmidt echo singularities in the exact dynamics, as well as their quantitative understanding via a model of 2LS as well as three-level systems (3LS); section~\ref{s.imbdynamics} discusses the dynamics of imbalance, and the comparison with the prediction of the 2LS and 3LS model; section~\ref{s.dephasing} discusses the departure of the exact data (for the Loschmidt echo and imbalance dynamics) from the 2LS/3LS predictions as a signature of $l$-bit interactions; section~\ref{s.entropy} shows that the 2LS/3LS models capture quantitatively the entanglement dynamics at short time, and of number entropy at longer times;  conclusions are drawn in section~\ref{s.conclusions}. 

\section{Model}
\label{s.Model}
Our platform for the investigation of Loschmidt-echo dynamics is given by a paradigmatic model, namely the $S=1/2$ XXZ chain in an inhomogeneous magnetic field \cite{Znidaric2008, Pal2010a}, corresponding to a model of spinless fermions with nearest-neighbor interactions in an inhomogeneous local chemical potential \cite{Oganesyan2007a}
\begin{equation}
\eqalign{
{\cal H} & = \sum_{i=1}^{L-1} \left [ -\frac{J}{2} \left ( S_i^+ S_{i+1}^- + {\rm h.c.} \right ) + J_z S_i^z S_{i+1}^z\right  ] - \sum_{i=1}^L h_i S_i^z \nonumber \\
& = \sum_{i=1}^{L-1} \left [ -\frac{J}{2}\left (c_i^\dagger c_{i+1} + {\rm h.c.} \right ) + J_z n_i n_{i+1} \right ] - \sum_{i=1}^L h_i n_i,
}
\end{equation}
where $S_i^\alpha$ ($\alpha = x,y,z$) are spin operators and $c_i, c_i^\dagger$ and $n_i = c_i^\dagger c_i$ are fermionic operators; the equality  between the two Hamiltonians is true up to an additive constant via Jordan-Wigner mapping. In the following the external field/potential $h_i$ is taken to be either quasi-periodic (QP)~\cite{Schuster2002, Iyer2013a}, namely $h_i = \Delta \cos(2 \pi \kappa i + \phi)$ with $\kappa = 0.721$ (inspired by experiments on bichromatic optical lattices \cite{Schreiberetal2015, Kohlert2019}) and $\phi$ a random phase; or to be fully random (FR) and uniformly distributed in the interval $[-\Delta, \Delta]$.  We consider chains of length $L$ (up to $L=22$) with open boundaries, and we average our results over $\sim 10^3$ realizations of the random phase (QP) or of the full random potential (FR). All the unitary evolutions considered in this study are obtained using exact diagonalization (ED), and they start from the charge-density wave state $|\psi_0\rangle = |1010101...\rangle$, corresponding to a N\'eel state for the spins.
We shall focus on the case of interacting fermions $J_z = J$ (corresponding to an SU(2) invariant spin-spin interaction) and contrast it with the limit of free fermions $J_z = 0$. In the latter case, the QP potential leads to a transition to fully localized single-particle eigenstates for $\Delta \geq J$, with an energy independent localization length $\xi = 1/\log(\Delta/J)$; while the FR potential leads to AL of the whole spectrum at any infinitesimal value of disorder. In the interacting case, instead, a QP potential of strength $\Delta \gtrsim 4J$  \cite{Naldesi2016} and a FR potential of strength $\Delta \gtrsim  3.5 J$ \cite{Luitz2015a, Singh2015, Khemani2017, Doggen2018, Doggen2019, Weiner2019, Sierant2020, Laflorencie2020} are numerically found to lead to MBL. In the following sections we shall generally start our discussion from the case of the QP potential, which has a simpler spatial structure devoid of rare regions, leading to stronger localization effects; and we shall later discuss how to enrich the picture in the case of the FR potential, in order to account for the existence of rare regions.

\section{Loschmidt echo singularities and imbalance oscillations}
\label{sub:LESING}
\begin{figure}[tb]
\centering
\includegraphics[width=0.6\textwidth]{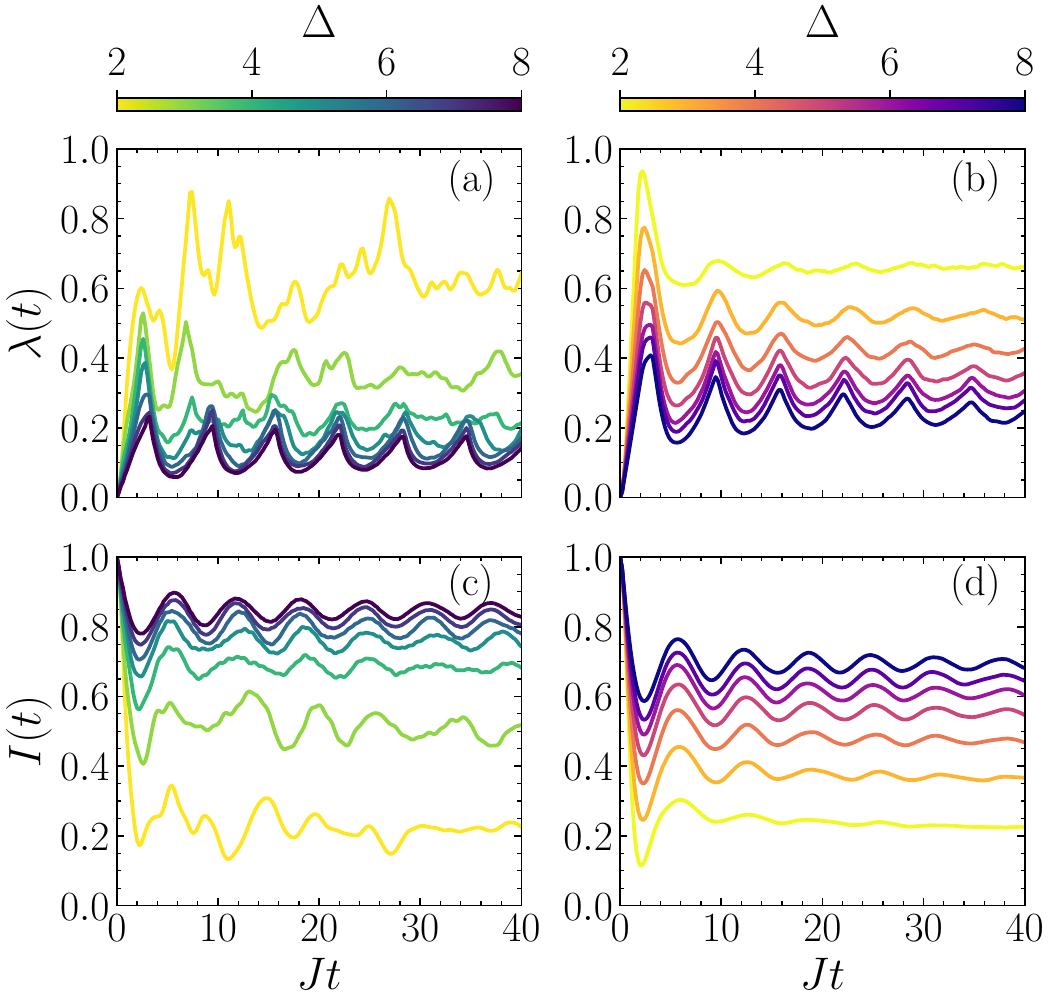}
\caption{Loschmidt echo and imbalance dynamics for an $L=22$ chain with QP potential (a-c) and FR potential (b-d), for various disorder strengths $\Delta = 2, \ldots, 8$ as indicated by the colors.}
\label{fig:Fig1}
\end{figure}
Figure~\ref{fig:Fig1} shows the dynamics of the Loschmidt echo $\lambda(t)$, ~\eref{e.LE},  along with that of the imbalance 
\begin{equation}
I(t) = \frac{1}{L}\sum_i (-1)^i (2[\langle n_i \rangle]_{\rm av}-1)~.
\end{equation}
 The latter saturates to its maximum value of 1 in the initial state and probes the persistence of the initial density/spin pattern \cite{Schreiberetal2015, Bordia2015, Luschenetal2017}. We observe that for both the QP and FR potentials, and for disorder strengths compatible with the onset of the MBL regime, the Loschmidt echo displays a sequence of periodic cusp-like peaks at times $t_m = (2m+1) \pi/J$ $(m=0, 1, 2, ...)$. These times correspond to minima in the imbalance, as the system reaches instantaneous configurations which are the farthest from the initial spin/density pattern.  

 A closer inspection shows that, for sufficiently strong disorder, all the peaks become sharp cusps, namely they represent genuine non-analyticities of the Loschmidt echo. They are rather remarkable given that they survive disorder averaging, and they seemingly appear in a finite fraction of disorder realizations (see \ref{App:LEreal} for further details); and in particular they decay very slowly in time, as we shall discuss in detail later on. The rest of this work will be devoted to developing a quantitative understanding of the dynamics of Loschmidt-echo singularities and imbalance oscillations as signatures of the existence of $l$-bits and of their interacting nature.

\section{Quantitative modeling of the Loschmidt echo singularities}

\subsection{Modeling with an ensemble of two-level systems}
\label{s.2LS}
\begin{figure}[b]
    \centering
    \includegraphics[width=0.6\columnwidth]{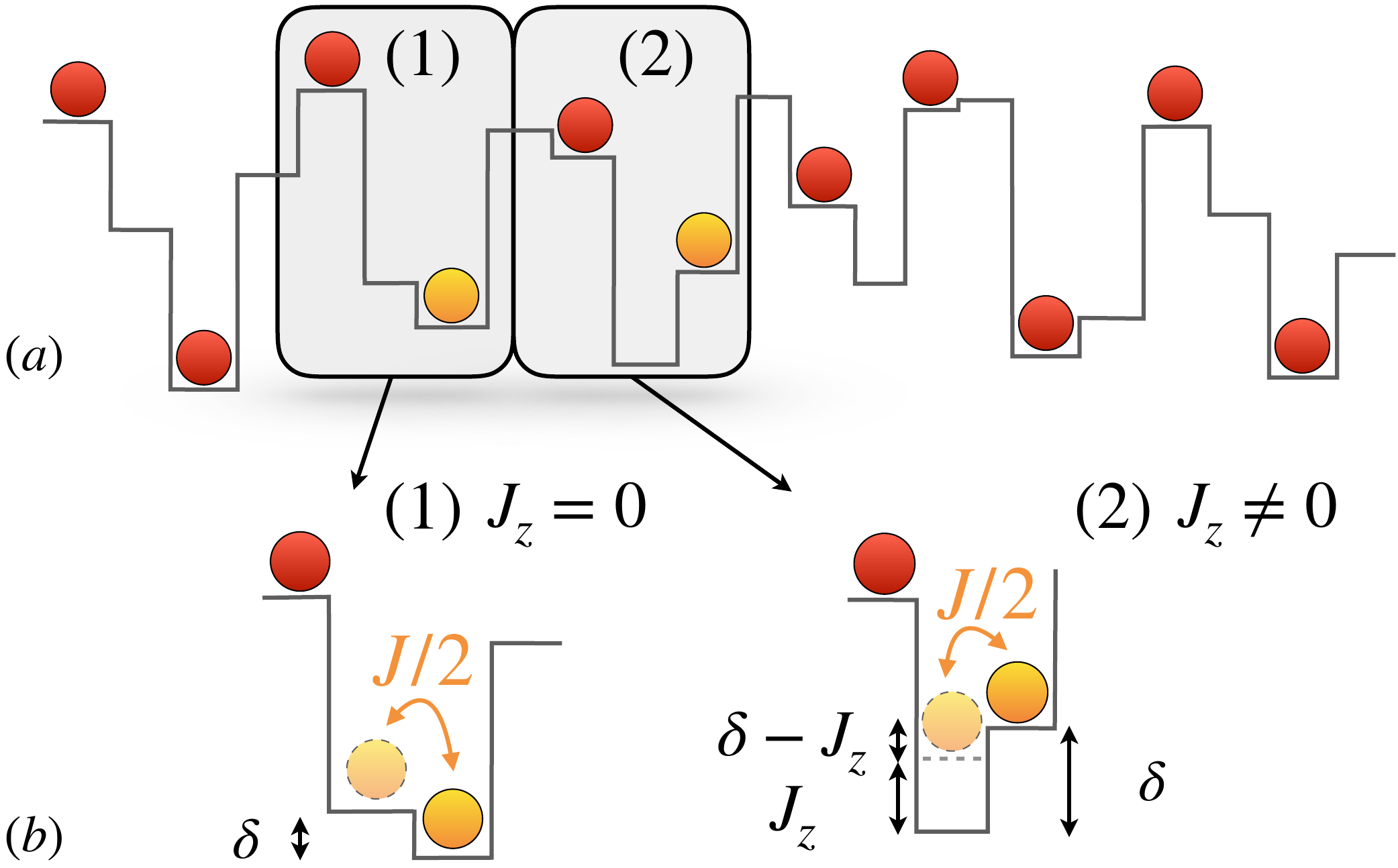}
    \caption{(a) Example of a $L=22$ chain in a QP potential (lines) in the initial CDW state $\ket{1 0 1 0...}$. Particles are denoted as coloured balls.
    (b) Zoom on two quasi-resonant regions (shaded areas): in the case of non-interacting particles $(J_z=0)$ the region (1) presents a pair of quasi-resonant sites for the particle in orange; in the case of interacting particles, region (2) shows two quasi-resonant sites for the orange particle, thanks to the partial screening of disorder offered by the interaction with the red particle.}
    \label{fig:Fig2}
\end{figure}

All the essential details of the short-time evolution of the Loschmidt echo can be captured with a surprisingly simple, yet rather insightful model.  This model is best understood (and justified) in the case of the QP potential, as illustrated in figure~\ref{fig:Fig2}. 

In the case of strong disorder, the fastest dynamics in the system starting from a Fock state will be offered by those particles that sit on a site $i$ which is nearly resonant with its unoccupied neighbor (say $i+1$), because the hopping $J/2$ is either larger than the energy offset $\delta_i = h_{i+1}-h_{i}$ (in the non-interacting case) or larger than the screened offset $\delta_i - J_z$ (in the presence of nearest-neighbor repulsion). These 2-site clusters, representing nearly resonant two-level systems (2LS), have the property of being spatially isolated in the QP potential, because of the strong anticorrelation among two consecutive energy offsets ($\delta_i$ and $\delta_{i\pm1}$ -- see \ref{App:SpatCorr}). As a consequence, a nearly resonant 2-site system will be generally surrounded by highly non-resonant pairs of sites, which can be considered as nearly frozen to the initial state. This invites us to write for the evolved state a 2LS Ansatz 
\begin{equation}
|\psi(t) \rangle \approx \left( \otimes_{p} |\psi^{(p)}_{2LS}(t)\rangle\right) \otimes \left( \otimes'_i |\psi_{0,i}\rangle \right), 
\end{equation}
where the first tensor product $\otimes_p$ runs over the nearly resonant 2LS, while the second tensor product $\otimes_i'$ runs over the leftover sites (we have taken the freedom of reordering the sites arbitrarily in the tensor product; see \cite{Sierant2017a, Sierant2017b, Janarek2018} for a similar Ansatz to study the long-time dynamics).  $|\psi^{(p)}_{2LS}(t)\rangle$ is the evolved state of the $p$-th (isolated) 2LS system, corresponding to two states split by an energy difference $\delta'_p = \delta_p-J_z$ and connected by a Rabi coupling $J$; while $|\psi_{0,i}\rangle$ is the (persistent) initial state of the site $i$ belonging to the remainder of the system.
The Loschmidt echo for such a system is readily calculated as
\begin{equation}
  \lambda(t) = -\frac{1}{L}\sum_p \log \left [ 1- p(\delta'_p,J,t) \right], \label{e.Rabi1}
 \end{equation}
with $p(\delta,\Omega,t) = (\Omega/\Omega')^2 \sin^2(\Omega' t/2)$ (and $\Omega' = \sqrt{{\Omega}^2 + \delta^2}$) the well-known probability of finding the 2LS in the state orthogonal to the initial one while performing Rabi oscillations \cite{QO-book} -- see also \ref{App:Rabi}. When averaging Eq.~\eref{e.Rabi1} over disorder, it is immediate to obtain the following simple expression
\begin{equation}
  \lambda_{2LS}(t) = -\int P(\delta'+J_z) \log \left [ 1- p(\delta';J,t) \right]\label{e.Rabi2},
\end{equation}
where $P(x)$ is the probability that the energy offset between two neighboring sites takes the value $x$. Going from Eq.~\eref{e.Rabi1} to Eq.~\eref{e.Rabi2} implies that we in fact count all of the $L$ pairs of sites in a chain as nearly resonant 2LS, thereby counting twice every site. The mistake that one makes in doing this is minor, though, because the non-resonant pairs of sites give a very small contribution to the Loschmidt echo; and, if neighboring pairs of sites are not simultaneously resonant, a site will not be counted twice in practice. Eq.~\eref{e.Rabi2} is an analytical integral formula which depends uniquely on the (known) statistics of the disorder potential via the $P$ distribution. In the case of the QP potential 
\begin{equation}
P(x) =  \frac{[1-(x/\tilde{\Delta})^2]^{-1/2}}{\pi\tilde{\Delta}}
\end{equation}
with $\tilde{\Delta} = \Delta \sin(\pi \kappa)$ \cite{Guarrera2007}; while for the FR potential $P(x)$
is the normalized  triangular distribution defined on the $[-2\Delta,2\Delta]$ interval. 

\begin{figure}[b]
    \centering
    \includegraphics[width=0.6\columnwidth]{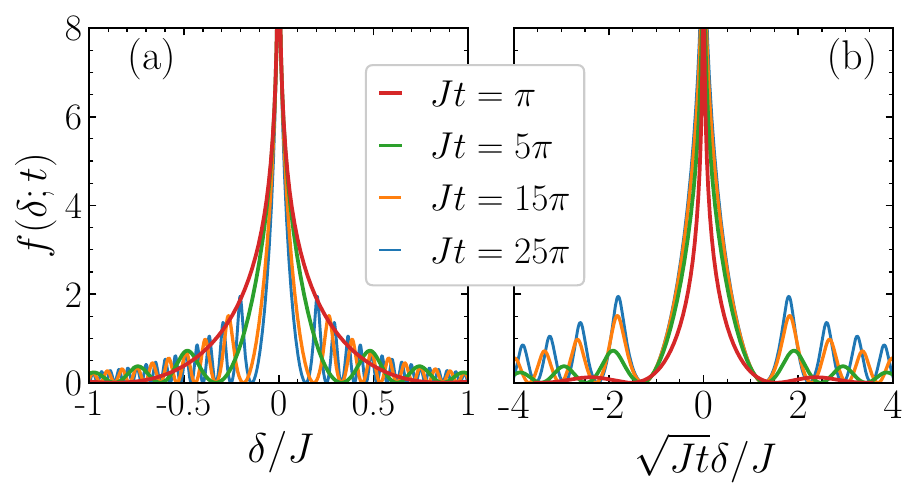}
    \caption{Function $f(\delta;t)$ vs. $\delta$ at different singularity times $t_m J = (2m+1)\pi$; as shown in the right panel, for large $t_m$ the width of the central peak becomes time-independent when $f$ is plotted as a function of $\sqrt{t}~\delta$.}
    \label{fig:fdelta}
\end{figure}

\begin{figure}[tb]
    \centering
    \includegraphics[width=0.6\columnwidth]{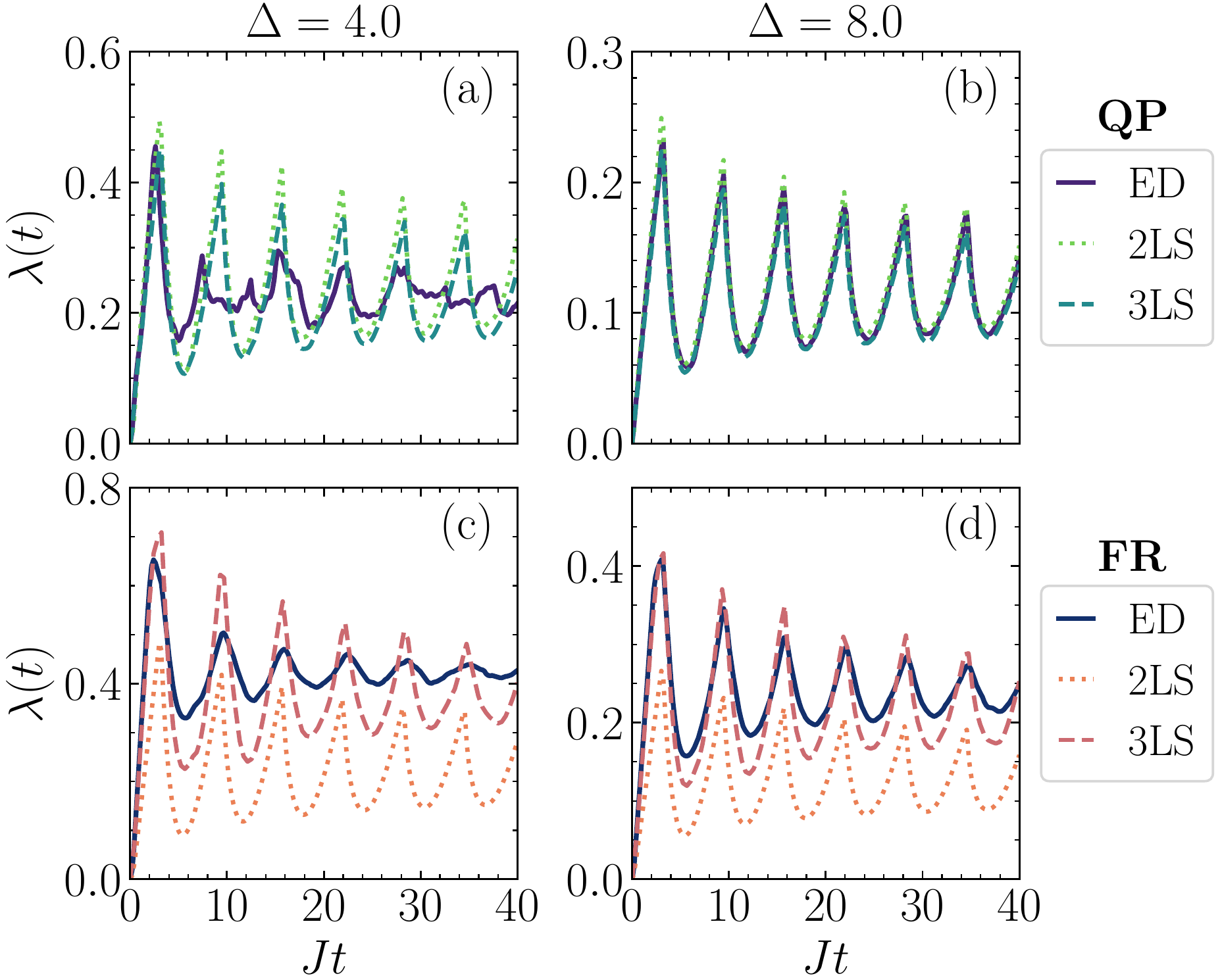}
    \caption{Comparison between the LE $\lambda(t)$ for and $L=22$ chain and the predictions of the 2LS and 3LS models: (a-c) QP potential; (d-f) FR potential.}
    \label{fig:Fig3}
\end{figure}

The cusp singularities in $\lambda(t)$ at times $t_m J= (2m+1)\pi$, $m = 0, 1, 2, \ldots$, descend from the fact that the integrand function $f(\delta;t_m)=\log[1-p(\delta, J, t_m)]$, seen as a function of $\delta$, develops a logarithmic singularity at $\delta=0$, as shown in figure~\ref{fig:fdelta}, while it is fully regular at any other time. The singular peak centered at $\delta=0$ has a support shrinking with $t_m$ as $t_m^{-1/2}$ -- as seen in figure~\ref{fig:fdelta} when plotting the function $f(\delta;t_m)$ vs $\delta \sqrt{t_m}$, which leads to a collapse of the peak widths at different times (when $m \gg 1$). The integral of the $f$ function outside the peak contributes to the regular part of the Loschmidt echo, while the integral of the peak dictates fundamentally the height of the cusps above the regular background (estimated as the long-time average $\bar \lambda$), namely the quantity
$\lambda_P(t_m) = \lambda(t_m) - \bar \lambda$. The decay of the height of these cusps as $t_m^{-1/2}$ will be verified numerically in section~\ref{s.dephasing}. 
Figure~\ref{fig:Fig3}(a-b) shows that, for the case of the QP potential, Eq.~\eref{e.Rabi2} is able to predict with high accuracy the ED results deep in the MBL phase without \emph{any} adjustable parameter. In particular the cusp singularities of the ED results are easily explained as descending from the divergent singularity of the Loschmidt echo for a fully resonant 2LS with $\Omega = \Omega' = J$, reaching a state orthogonal to the initial one after odd multiples of half a Rabi oscillation $t_m = (2m+1) \pi/\Omega$. These divergences are smoothened into cusp singularities due to the fact that such resonant 2LSs are a set of zero measure in the disorder statistics. This result has important consequences. Indeed the nearly resonant 2LSs captured by the model are clearly an ensemble of approximate $l$-bits with Hamiltonian 
\begin{equation}
{\cal H} \approx \sum_p K_p \tau_p,
\label{eq:lbitHam}
\end{equation}
where $\tau_p = \delta'_p/K_p \sigma^z_p - J/K_p \sigma_p^x$ is a Pauli matrix expressed as a rotation of the Pauli operators $\sigma^z_p = S^z_{i+1} - S^z_i$ (when projected onto the subspace with $S_i^z+S_{i+1}^z = 0$) and $\sigma^x = S_i^+ S_{i+1}^- + {\rm h.c.}$, built from the original spin operators for the pair $p=(i,i+1)$; and $K_p = \sqrt{\delta_p^2 + J^2}$ is the $l$-bit splitting. Hence the Loschmidt-echo singularities are a striking manifestation of the existence of such (nearly free) $l$-bits, to be found in the short-time dynamics of the system.

\subsection{Relationship to dynamical quantum phase transitions}

It is worthwhile to mention at this point that the existence of singularities in the quench dynamics of the Loschmidt echo is currently the subject of several theoretical and experimental investigations, as they represent the main signature of so-called dynamical quantum phase transitions, studied both in non-random systems \cite{Heyl2013, Heyl2014,Jurcevicetal2017,Heyl2018, Guoetal2019} as well as in disordered quantum systems \cite{Yang2017, Yin2018, Halimeh2019}. Nonetheless our observation of Loschmidt-echo singularities is fully explained by a model of individual 2LS, without the need of any many-body effect. Therefore we shall refrain from associating them to any form of time-dependent transition.


\subsection{From two-level systems to three-level ones}
\label{sub:3LS}
Figure~\ref{fig:Fig3}(d-f) shows that, in the case of the FR potential, the 2LS model of ~\eref{e.Rabi2} still predicts the correct frequency of the Loschmidt echo singularities, but not the correct height; and it also misses a global offset. This is not surprising, as in the case of the FR potential the assumption of anti-correlation between the energy offset of contiguous pairs is no longer valid, namely the potential can host ``rare" regions in which contiguous pairs of sites -- $(i-1,i)$ and $(i,i+1)$ -- are nearly resonant at the same time. To take those regions into account (at least partially) one can easily promote the 2LS model to a model of 3-site systems (amounting to effective three-level systems -- 3LS), and approximate the evolved state as that of a collection of independent 3LS. The Hamiltonian of a three-site system $(i,i+1,i+2)$ containing two interacting fermions in an initial $\ket{101}$ state is explicitly given by
\begin{equation}
\eqalign{
    {\cal H}_{101} & = -\frac{J}{2}\left(c^\dagger_i c_{i+1} + c^\dagger_{i+1} c_{i+2} + {\rm h.c.} \right) \\
    & + \delta_i n_{i+1} + (\delta_{i+1}+\delta_i) n_{i+2} \\
    & + J_z \left (n_i n_{i+1} + n_{i+1} n_{i+2} \right) +{\rm const.}
    }
\label{eq:Ham101}
\end{equation}
Here all single-site energies are referred to the energy of site $i$, and $\delta_{i} = h_{i+1} - h_i$. The above Hamiltonian assumes that the sites $i-1$ and $i+3$ remain empty during the time evolution.  
The Hilbert space of the 3-site system is restricted to the three states $|101\rangle$, $|110\rangle$ and $|011\rangle$, making of it a three-level system (3LS), with a generic time-dependent wavefunction 
\begin{equation}
\ket{\psi_{101}(t)}=\alpha(t)\ket{011}+\beta(t)\ket{101}+\gamma(t)\ket{110}~.
\end{equation}
Its explicit form can be easily calculated numerically for any specific choice of the energy differences $\delta_i$. 

A similar calculation can be done for a three-site system hosting a single particle, and starting from the $|010\rangle$ configuration, with Hamiltonian 
\begin{equation}
\eqalign{
    {\cal H}_{010} & = -\frac{J}{2}\left(c^\dagger_i c_{i+1} + c^\dagger_{i+1} c_{i+2} + {\rm h.c.} \right) \\
    & + J_z n_i + \delta_{i} n_{i+1} + (\delta_{i+1} + \delta_i  + J_z) n_{i+2},
    }
\label{eq:Ham010}
\end{equation}
which assumes that the sites $i-1$ and $i+3$ host two pinned particles. The Hilbert space $|100\rangle$, $|010\rangle$, $|001\rangle$ defines a 3LS, whose instantaneous state takes the generic form
\begin{equation}
\ket{\psi_{010}(t)}=\tilde\alpha(t)\ket{100}+\tilde\beta(t)\ket{010}+\tilde\gamma(t)\ket{001}~.
\end{equation}
For the two types of clusters the Loschmidt echo can be readily evaluated as $\lambda_{101}(t;\delta_i,\delta_{i+1})=-\log|{\beta}(t)|^2$ and $\lambda_{010}(t;\delta_i,\delta_{i+1})=-\log|{\tilde\beta}(t)|^2$.

 We can then model a chain in a QP or FR potential as an ensemble of independent 3LSs by generating sequences of energy offsets $\delta_i,\delta_{i+1}$ between adjacent site pairs according to the distribution $P(\delta_i,\delta_{i+1})$. The 3LS prediction for the Loschmidt echo of the ensemble is 
 \begin{equation}
 \lambda_{\rm 3LS}= \frac{1}{2} \int d\delta_1 d\delta_2 ~ P(\delta_1,\delta_2) 
 \left [ \lambda_{101}(t;\delta_1,\delta_2) + \lambda_{010}(t;\delta_1,\delta_2) \right ]~.
\end{equation}
In practice, the above integral can be sampled numerically by simply averaging over a large number of different realizations of the potential on 3-site systems, such as those offered by a very long chain, namely 
\begin{equation}
 \lambda_{\rm 3LS}(t) \approx \frac{1}{L} \sum_{i=1}^L ~\lambda_{\alpha_i}(t;\delta_i,\delta_{i+1}),
 \label{e.LE_3LS}
\end{equation}
where $\alpha_i = 101$ if $i$ is odd and  $010$ if $i$ is even, and $L\gg 1$.
 
Eq.~\eref{e.LE_3LS} for the Loschmidt echo has the apparent drawback of triple-counting each site. Nonetheless, similarly to what was argued for the 2LS case, it is fair to assume (and it can be numerically tested) that, out of the three clusters containing each site, only one at most will contribute significantly to the Loschmidt echo. As a consequence the triple counting has only a mild effect on the final result. One could avoid triple counting by thoughtfully decomposing a chain into non-overlapping clusters of up to 3 sites, in such a way as to maximize the Loschmidt echo; yet this procedure introduces significant complications which are not justified a posteriori, given the quality of the results offered already by the naive ensemble average. 
As shown by figure~\ref{fig:Fig3}(c-d), the improvement offered by the 3LS model for the FR potential is substantial; these results can further be improved by moving to 4-site clusters etc., albeit at an exponential cost.



\section{Imbalance dynamics} 
\label{s.imbdynamics}

 
The 2LS model prediction for the imbalance is very similar to that of the Loschmidt echo, as the imbalance is simply related to the persistence probability of the initial state ($|10\rangle$ or $|01\rangle$) on the 2-site cluster -- given that the orthogonal state contributes zero to the imbalance. 
Therefore the 2LS expression for the imbalance simply reads
\begin{equation}
 I_{2LS}(t) = \int d\delta ~P(\delta) ~\left [ 1 - p(\delta,t)\right]~.
 \label{eq:Imb2S}
\end{equation}
The times $t_m$ giving cusp singularities in the Loschmidt echo correspond to dips in the imbalance, and these dips come from local dips in the $g(\delta;t)=1-p(\delta,t)$ function centered around $\delta=0$ and touching zero for $t=t_m$. The width of these dips is also shrinking in time as $t_m^{-1/2}$. Therefore one expects the depth of the minima in the fluctuations of the imbalance to decay to the long-time average 
as $t_m^{-1/2}$ as well -- this prediction will be verified in section~\ref{s.dephasing}. 
\begin{figure}[b]
    \centering
    \includegraphics[width=0.6\columnwidth]{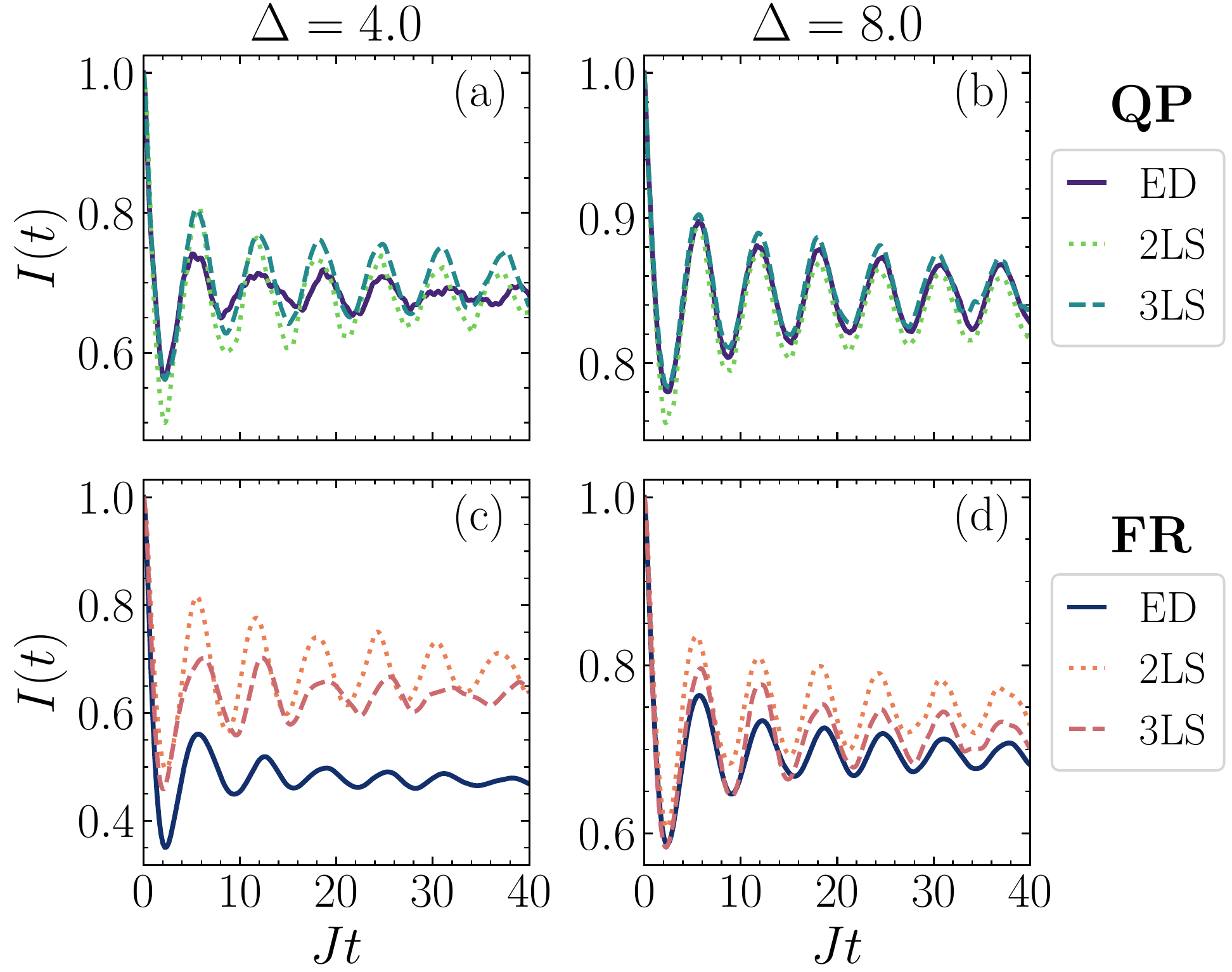} 
    \caption{Comparison between the imbalance $I(t)$ for a $L=22$ chain and the predictions of the 2LS and 3LS models given by \eref{eq:Imb2S}, \eref{eq:Imb3S}: (a-c) QP potential; (d-f) FR potential.}
   \label{fig:Imb}
\end{figure}

Extending the 2LS model to 3LS, the imbalance can be calculated as
  \begin{eqnarray}
I_{\rm 3LS}(t)  \approx \frac{1}{3L} \sum_i \left ( -|\alpha_i|^2 + 3|\beta_i|^2 - |\gamma_i|^2\right )~~~
\label{eq:Imb3S}
\end{eqnarray}
with $\alpha_i,\beta_i,\gamma_i = \alpha(t),\beta(t),\gamma(t)$ or $\tilde \alpha(t), \tilde \beta(t), \tilde \gamma(t)$ depending on whether $i$ is odd or even.
Figure~\ref{fig:Imb} shows the comparison between the ED results for the imbalance dynamics of interacting fermions immersed in a QP and fully random potentials of variable strength, compared with the predictions of the 2LS and the 3LS models. For sufficiently strong disorder ($\Delta \gtrsim 6 J$) the 2LS predictions are already rather accurate in the case of the QP potential, and the 3LS model offers further improvement. On the other hand in the case of the FR potential the 3LS model offers a more substantial improvement, fixing an overall offset (for sufficiently strong disorder) which is seen in the 2LS predictions. 
\section{Dephasing in the Loschmidt echo and imbalance oscillations: evidence of \texorpdfstring{$l$}{l}-bits interactions}
\label{s.dephasing}
A significant feature of the Loschmidt echo singularities is their slow decay in time -- which is remarkable given that they result from the Rabi oscillations of a collection of 2LS with a distribution of frequencies that can be \emph{a priori} expected to lead to fast dephasing. The reason behind the slow decay is also captured by the 2LS model, Eq.~\eref{e.Rabi2} -- namely by the fact that the integral expressing the Loschmidt echo takes contributions from a small window of detunings $\delta'$ around zero, the smaller the longer the time, as mentioned in section~\ref{s.2LS}. When looking at the singularity times $t=t_m$, a direct inspection of the function $\log(1-p(\delta',\Omega,t_m))$ seen as a function of $\delta'$ shows that it has a large peak centered on $\delta'=0$ with a width depending on time as $t_m^{-1/2}$. The singularity in the average Loschmidt echo comes from the integral of this peak, while the rest of the integral contributes essentially to the regular part of the Loschmidt echo; hence it is immediate to predict that the height of the cusp singularity should decay as the peak width, (namely as $t_m^{-1/2}$).

\begin{figure}[t]
    \centering
    \includegraphics[width=0.6\columnwidth]{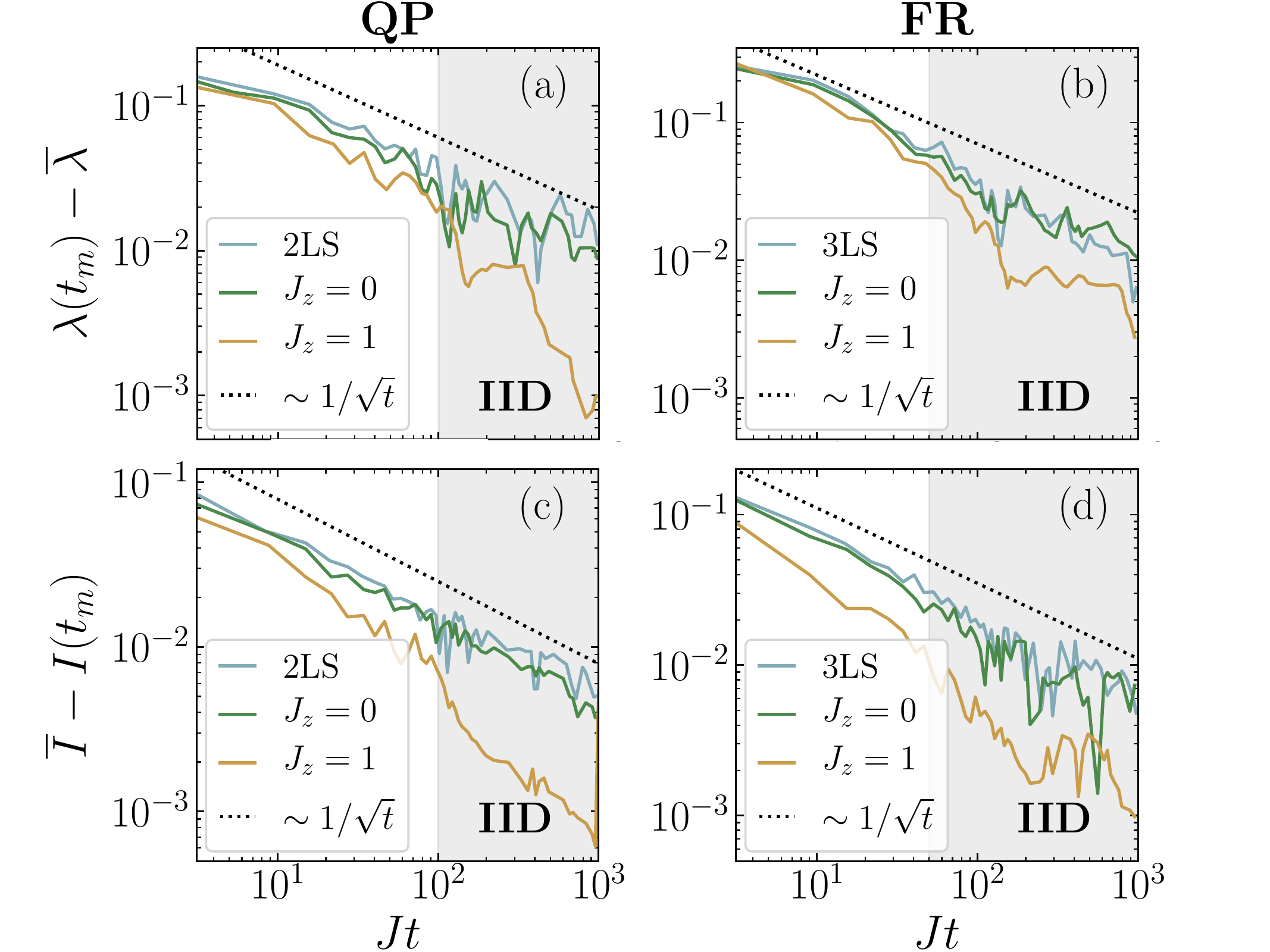}
    \caption{(a-b) Decay of the peak heights of the Loschmidt echo, $\lambda(t_m) - \bar\lambda$ ($\bar\lambda$ stands for the time-averaged Loschmidt echo).  (c-d) Decay of the depth of the imbalance minima $I(t_m)$ with respect to the average value $\overline{I}$. ED data in both absence ($J_z=0$) and presence ($J_z=1$) of interactions are compared with 2LS and 3LS predictions. The data are obtained for $\Delta=8J$; the 2LS and 3LS predictions are for $J_z = J$. The grey-shaded area marks the interaction-induced dephasing (IID) regime exhibited by the exact data for $J_z=J$.}
    \label{fig:Fig4}
\end{figure}

Figure~\ref{fig:Fig4}(a) shows the time evolution of singularity peaks in the Loschmidt echo for free as well as interacting fermions in the QP potential, compared to the prediction of the 2LS model (for the interacting case): we observe that the $t^{-1/2}$ decay is indeed confirmed by the ED data for free fermions, as well as by the ED data for interacting fermions at sufficiently short times ($tJ \lesssim t^* \approx 100$ for $\Delta = 8J$).  On the other hand, at longer times the interacting data are found to display a strong deviation from the 2LS model prediction, exhibiting a much faster decay. This crossover to an interaction-induced dephasing (IID) regime clearly shows the limits of the 2LS model as a model of free $l$-bits, and it marks a fundamental difference between AL and MBL in the QP system. Indeed the faster decay of the Loschmidt echo must be related to the effect of $l$-bit interactions, which are a defining feature of MBL, and which add terms of the kind 
\begin{equation}
\sum_{pq} U_{pq} \tau_p \tau_q + \sum_{pql}  V_{pql} \tau_p \tau_q \tau_l + ...
\end{equation}
to the effective $l$-bit Hamiltonian \eref{eq:lbitHam}. Such terms are responsible for the persistent growth of entanglement entropy in the system \cite{Serbynetal2013} as the logarithm of time, and indeed the onset of the $\log t$ growth of entanglement occurs at a time compatible with $t^*$ (see section~\ref{s.entropy} - figure~\ref{fig:entQP}).  A similar crossover from a slow power-law decay of the Loschmidt-echo peak height to a faster decay, dictated by the presence of interactions, is also exhibited by the comparison between the ED data for interacting fermions in the FR potential with the same data for non-interacting fermions and for the 3LS model -- as shown in figure~\ref{fig:Fig4}(b). 

Remarkably, the same crossover between the dynamics of effectively independent $l$-bits to a regime of interacting ones is observed in the decay of imbalance oscillations. 
Figure~\ref{fig:Fig4}(c) shows the evolution of the depth of the minima of the imbalance at times $t=t_m$, taken with respect to the long-time average, namely the quantity $I_M(t_m) = \bar{I} - I(t_m)$. We observe that the predictions of the 2LS system for the fermionic chain immersed in the QP potential shows a clear, slow power-law decay at long times, compatible with $t^{-1/2}$, which is indeed reproduced in the case of non-interacting fermions. In the case of interacting fermions, on the other hand, a crossover is observed at long times ($t \gtrsim t^* \approx 100$) to a faster decay, marking the IID regime. A similar picture is offered by the case of the FR potential. There the ED results are compared with the predictions from the 3LS model; the latter model predicts correctly the decay of the minima depth in the non-interacting case at all times, while the exact results for the interacting system show a clear crossover towards a faster decay for times  $t \gtrsim t^* \approx 50$.
For both kinds of disorder, the crossover time $t^*$ is compatible both with what is observed in the decay of the Loschmidt echo as well as with the evolution of the entanglement entropy (see again figure~\ref{fig:entQP} in the next section). Therefore we conclude that the crossover to the IID regime is a robust feature of MBL dynamics, clearly exposing the interactions among $l$-bits.

\section{Entanglement dynamics}
\label{s.entropy}

\subsection{Entanglement entropy from the 2LS and 3LS model}

The 2LS and 3LS models allow for a simple calculation of the entanglement entropy of a $A$/$B$ bipartition of the system into two adjacent chains, defined as the von Neumann entropy of the reduced density matrix
\begin{equation}
 S_A(t)= - {\rm Tr} \left[ \rho_A(t) \log \rho_A(t) \right],
\end{equation}
where $\rho_A(t) = {\rm Tr}_B |\psi(t)\rangle \langle \psi(t)|$ is the partial trace (over the degrees of freedom in $B$) of the instantaneous pure-state density matrix associated with the evolved state $|\psi(t)\rangle$.
  
The entanglement associated with such a bipartition simply comes from the entanglement inside the 2-site or 3-site cluster which contains the cut defining the bipartition. In the case of the 2LS model the disorder-averaged entanglement entropy of a bipartition is simply predicted as the entropy of the reduced state of one site in the 2-site cluster, namely
\begin{equation}
S_A(t) =  \int ~d\delta~P(\delta) ~ h[p(\delta;t)],
\end{equation}
and $h[x] = -x\log x - (1-x) \log (1-x)$.  Notice that, unlike for the formulas of the Loschmidt echo and of the imbalance, no-double counting is implied in the above formula, since the entanglement is referred to a cut of the chain, and there is one unique cut per 2-site cluster.    
  
On the other hand a 3-site cluster can be cut in two different ways, that we will indicate as ${\rm \circ~|\circ\circ}$ and ${\rm \circ\circ|~\circ}$ in the following (where $\circ$ stands for a site and $|$ stands for the cut). The reduced density matrices for the two cuts are readily obtained from the cluster wavefunctions  described in section~\ref{sub:3LS}; e.g. for a 101 cluster the reduced density matrix associated with the ${\rm \circ~|\circ\circ}$ cut reads   
\begin{equation}
    \rho_{101}^{\rm \circ|\circ\circ}(t) = 
    \left( 
    \matrix{|{\alpha}(t)|^2 & 0 \cr
    0 & |{\beta}(t)|^2+|{\gamma}(t)|^2}
    \right),
\label{eq:rhox|xx}
\end{equation}
with associated entanglement entropy 
\begin{equation}
S_{101}^{\rm \circ|\circ\circ}= -  |{\alpha}(t)|^2 \log  |{\alpha}(t)|^2
- \left ( |{\beta}(t)|^2+|{\gamma}(t)|^2 \right ) \log \left ( |{\beta}(t)|^2+|{\gamma}(t)|^2 \right);
\end{equation}
the one associated with the  $\circ\circ|~\circ$ cut reads
\begin{equation}
    \rho_{101}^{\rm \circ\circ | \circ}(t) = 
    \left( 
    \matrix{
    |{\gamma}(t)|^2 & 0 \cr
    0 & |{\alpha}(t)|^2+|{\beta}(t)|^2}
    \right).
\label{eq:rhoxx|x}
\end{equation}
The density matrices $\rho_{010}^{\rm \circ|\circ\circ}$ and $\rho_{010}^{\rm \circ\circ | \circ}$ and related entropies associated with a 010 cluster can be calculated similarly. 
The disorder-averaged entanglement entropy of the whole system within the 3LS model is then given by
\begin{equation}
    \mathcal{S}(t)=\frac{1}{2L}\sum_{i}S_{i}(t),
    \label{eq:ent3LS}
\end{equation}   
where   
\begin{equation}
      S_i=
    \cases{
    S_{101}^{\rm \circ|\circ\circ}(t)+S_{101}^{\rm \circ\circ | \circ}(t) & if \textit{i} odd,\\
    S_{010}^{\rm \circ|\circ\circ}(t)+S_{010}^{\rm \circ\circ | \circ}(t) & if \textit{i} even.
    }
\end{equation}
The factor $1/2$ in \eref{eq:ent3LS} comes from the double counting of each cut (which is contained both in a 101 cluster as well as in a 010 cluster).

\begin{figure}[tb]
    \centering
    \includegraphics[width=0.6\columnwidth]{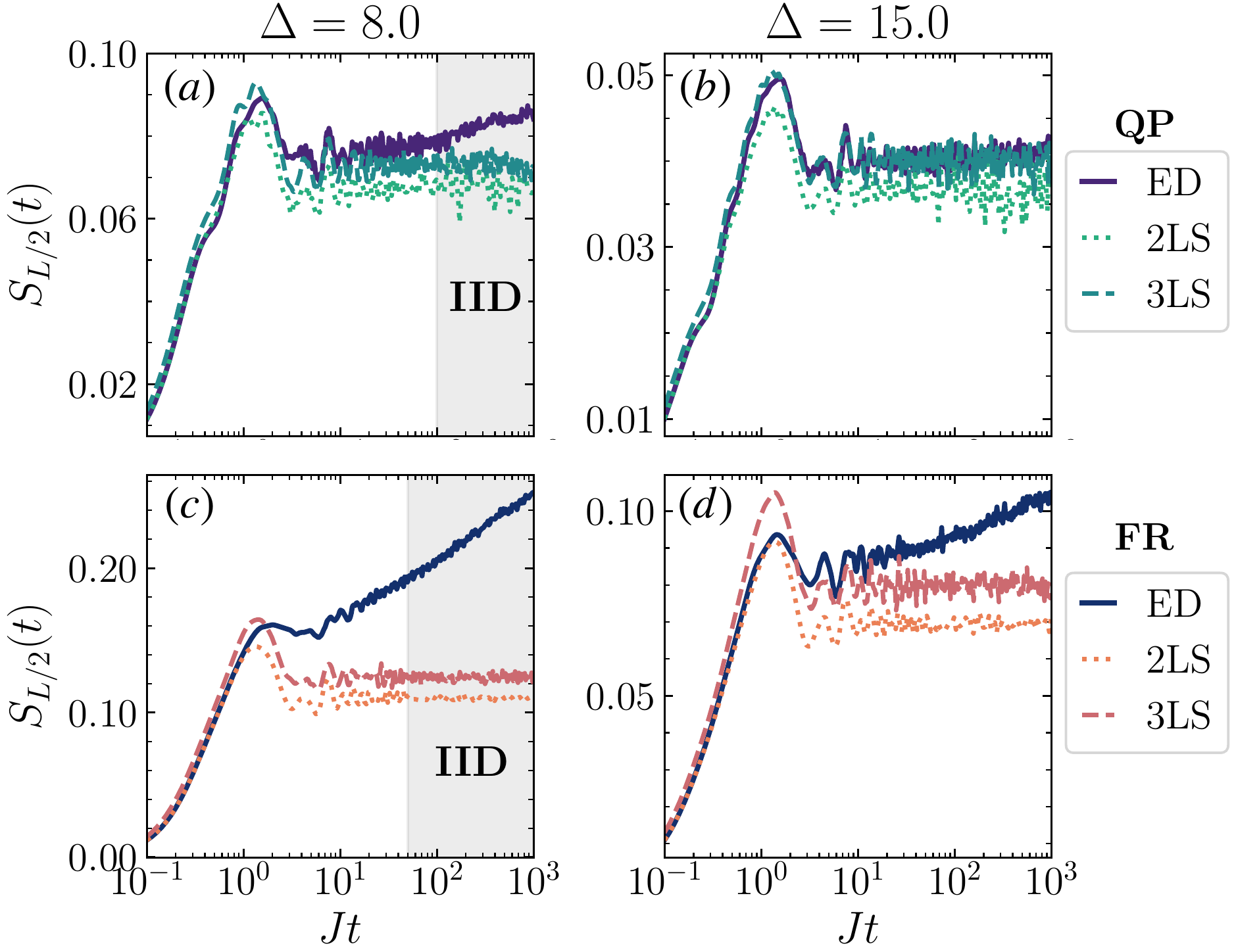}
    \caption{Half-chain entanglement entropy of interacting fermions in a QP potential (a-b) and FR potential (c-d) for a chain of size $L=16$, compared with the prediction for the 2LS/3LS models for two different disorder strengths ($\Delta/J = 8$ and 15).}
    \label{fig:entQP}
\end{figure}

Figure~\ref{fig:entQP} shows a comparison between the entanglement entropy of interacting fermions in a QP potential and the 2LS prediction. We observe that at moderate disorder in the MBL phase ($\Delta = 8 J$) the 2LS and 3LS models only capture the initial rise of the entanglement entropy and (partly) the first maximum; in particular the very existence of a maximum is explained by the models as the result of nearly resonant small clusters returning close to the initially factorized state -- albeit at different times due to the inhomogeneously broadened local frequencies, which explains why the entanglement entropy does not come back to (nearly) zero. The 2LS and 3LS models on the other hand completely miss the long-time logarithmic growth of the entanglement entropy -- something which is fully expected, given that such a growth is the consequence of interactions between $l$-bits, not included in the 2LS and 3LS models by construction. 
On the other hand, at stronger disorder ($\Delta = 15 J$) the interactions between $l$-bits are parametrically suppressed, and the 2LS and 3LS description of entanglement becomes accurate up to very long times.  

\subsection{Entanglement entropy vs.\ number entropy}
\begin{figure}[t]
    \centering
    \includegraphics[width=\textwidth]{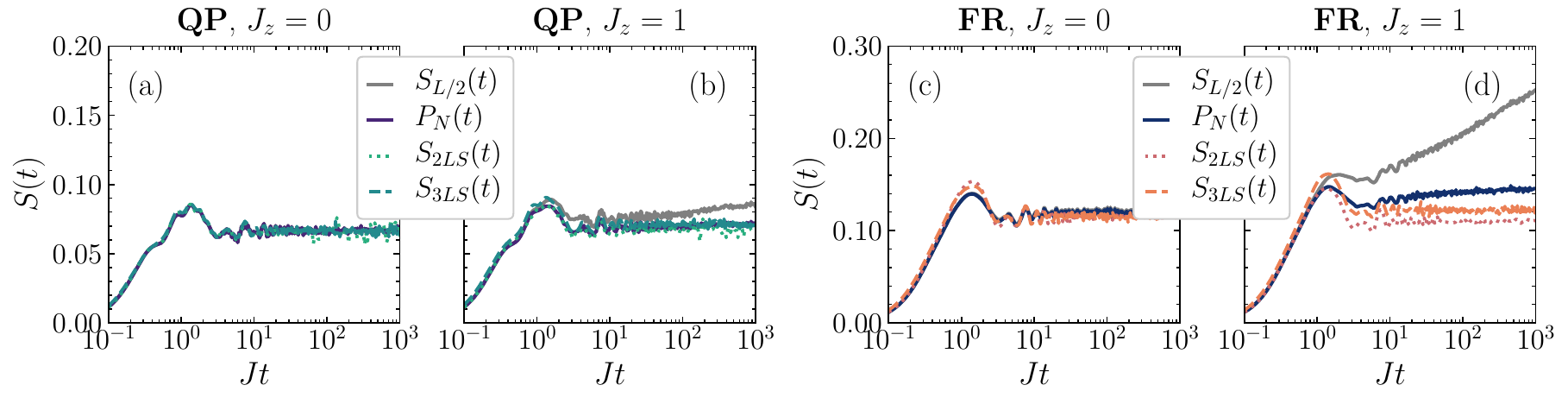}
    \caption{Half-chain entanglement entropy and number entropy of free and interacting fermions on a $L=16$ chain,  compared with the 2LS and 3LS predictions: (a) non-interacting fermions in a QP potential; (b) interacting fermions $(J_z=1)$ in a QP potential; (c) non-interacting fermions in a FR potential; (d) interacting fermions $(J_z=1)$ in a FR potential. For all the panels the disorder strength is $\Delta = 8 J$.}
    \label{fig:entvsnum}
\end{figure}

The 2LS and 3LS models picture the entanglement between two adjacent subsystems as arising uniquely from the coherent motion of particles within the restricted size of the clusters they describe. When starting from a factorized state, this picture is certainly valid at short times. At long times it remains valid only if particles remain localized within the size of the clusters (namely if the localization length is smaller than the cluster size), and if this is a sufficient condition for entanglement not to spread any further. The latter aspect is true in the case of non-interacting fermions, for which the only mechanism behind entanglement of different spatial partitions is particle motion between them. On the other hand, in the case of interacting fermions in the MBL regime, entanglement keeps growing due to the interactions between $l$-bits, and distant degrees of freedom can become entangled even without any net particle exchange. 
In this context it is useful to decompose the entanglement entropy of a subsystem $A$ into a number entropy contribution, and a remainder part (called the configurational entropy), $S_A = S_{A,N} + S_{A,c}$ \cite{Lukin2019,Kieferetal2020}. The number entropy is given by 
\begin{equation}
 S_{A,N} = - \sum_{N_A} p_{N_A} \log p_{N_A},
 \label{eq:numEntropy}
\end{equation} 
where $p_{N_A}$ is the probability of having $N_A$ particles in subsystem $A$. Eq.~\eref{eq:numEntropy} accounts for the particle number uncertainty appearing in subsystem $A$ because of the coherent exchange of particles with its complement $B$. On the other hand the configurational entropy accounts for correlations establishing between the particle arrangements in $A$ and $B$ once the partitioning of the particles between $A$ and $B$ has been fixed. 
The 2LS and 3LS models, completely lacking any form of correlations among the clusters, can only capture the number entropy contribution in systems with a localization length smaller than the cluster size. Nonetheless, this limited picture still offers a faithful description of entanglement in the MBL regime for short times (the longer the stronger disorder is), while it can describe entanglement at all times for strongly localized non-interacting particles.  
Thus, as suggested above, a more appropriate comparison with the entanglement entropies of the 2LS and 3LS models would involve the number entropy from the ED data -- shown in figure~\ref{fig:entvsnum}. For non-interacting fermions in a QP potential of strength $\Delta = 8 J$, $S_{A,N}$ is found to nearly coincide with the full entanglement entropy, and to be very well described by the 2LS prediction -- see figure~\ref{fig:entvsnum}(a). When adding the interactions, the agreement between the number entropy and the 2LS entropy deteriorates mostly at long times, seemingly due to the $\sim \log\log t$ growth of the number entropy observed in the MBL phase \cite{Kieferetal2020}. Similar considerations can be made in the case of the FR potential, i.e. the 3LS models describe well the entropies in the non-interacting case, and they miss the slow long-time growth of the number entropy in the interacting case. 

\section{Conclusions}
\label{s.conclusions}
In this work we have shown that sharp cusp-like singularities in the Loschmidt echo are a generic feature of the localized dynamics of an extended quantum system initialized in a factorized state. These features can be fully explained by the dynamics of a simple model, describing an ensemble of effective independent two-level (or even three-level) systems, offering an explicit approximation to the conserved $l$-bits in the AL and MBL regime. Such a model predicts very accurately the Loschmidt echo singularities for strongly disordered systems as well as their decay, along with the imbalance oscillations. A faster decay in the Loschmidt echo and imbalance dynamics compared to that predicted by the model is a direct manifestation of the dephasing effect of interactions between the $l$-bits, and it intervenes at a time consistent with the onset of the logarithmic growth of entanglement entropy: hence it represents a defining feature of many-body localization (MBL) with respect to Anderson localization (AL). Based on our results, we can conclude that experimental evidence of $l$-bit dynamics and of their interactions is readily accessible to state-of-the-art quantum simulators which have direct access to the Loschmidt echo and imbalance dynamics, such as \emph{e.g.} trapped ions \cite{Gaerttneretal2017, Jurcevicetal2017}, cold-atom simulators \cite{Schreiberetal2015, Luschenetal2017, Lukin2019} or superconducting circuits \cite{Roushanetal2017, Chiaroetal2020}.

\ack{L.B.\ gratefully acknowledges hospitality and financial support from the Laboratoire de Physique of the ENS Lyon. R.A.R.\ and L.B.\ acknowledge funding from the CY Initiative of Excellence (grant”Investissements d’Avenir” ANR-16-IDEX-0008) where this work developed during R.A.R.’s stay at the CY Advanced Studies. L.B.\ thanks the EPSRC for DTP funding. We thank Warwick's Scientific Computing Research Technology Platform and HPC Midlands+ (Athena) for computing time and support (EPSRC on grant EP/P020232/1). Part of the exact diagonalization simulations were performed using routines contained in the \emph{QuSpin}~\cite{Weinberg2017QuSpin:Chains, Weinberg2018} and \emph{quimb}~\cite{gray2018quimb} open-source libraries.}

\section*{References}
\bibliographystyle{iopart-num}
\bibliography{references.bib}

\appendix

\section{Two-site cluster as a two-level system and its Rabi oscillations}
\label{App:Rabi}
Let us isolate a two-site system $(i,i+1)$ hosting one particle in the fermionic chain, with Hamiltonian 
\begin{equation}
{\cal H}_{\rm 2-site} = -\frac{J}{2} \left ( c^\dagger_i c_{i+1} + c^\dagger_{i+1} c_i \right )  + h_i n_i + (h_{i+1}+J_z) n_{i+1},    
\end{equation}
where we assume that the site $i+2$ is occupied by a (pinned) particle, while size $i-1$ is empty (or occupied by a pinned hole). 
Introducing the spin operators 
\begin{eqnarray}
\sigma^z & = &  n_i - n_{i+1}, \nonumber \\ 
\sigma^x & = &  c^\dagger_i c_{i+1} + c^\dagger_{i+1}, c_i, 
\end{eqnarray}  
the Hamiltonian becomes simply
\begin{equation}
{\cal H}_{\rm 2-site} = -\frac{J}{2} \sigma^x + \frac{\delta}{2} \sigma^z + {\rm const.},
\end{equation}
namely, a two-level system (2LS) with splitting $\delta$ and Rabi frequency $J$. If the system starts from the $|10\rangle$ state, the return probability is given by the well-known formula for the probability of persistence in the initial state during Rabi oscillations \cite{QO-book}, namely $1-p(\delta;t)$, where
\begin{equation}
 p(\delta;t) = \frac{1}{1+(\delta/J)^2} ~\sin^2 \left ( \frac{\sqrt{1+(\delta/J)^2}}{2} tJ \right ) ~.
 \label{e.p}
\end{equation}

\section{Spatial correlations in the quasi-periodic vs. fully random potential}
\label{App:SpatCorr}
A fundamental assumption of the 2LS model described in the main text is that quasi-resonant two-site systems are spatially isolated in a (quasi)-disordered chain -- namely,  if a pair of sites $(i,i+1)$ is quasi-resonant for the motion of a particle, the two adjacent pairs of sites $(i-1,i)$ and $(i+1,i+2)$ are not resonant. Defining $\delta_1 = h_{i+1}-h_i$ as the energy difference of the two sites in question, and $\delta_2 = h_{i+2}-h_{i+1}$ as that of the following pairs of sites, in the case of non-interacting fermions, the above condition requires that the two energy differences do not vanish simultaneously. 
 
\begin{figure}[tb]
    \centering
    \includegraphics[width=0.7\textwidth]{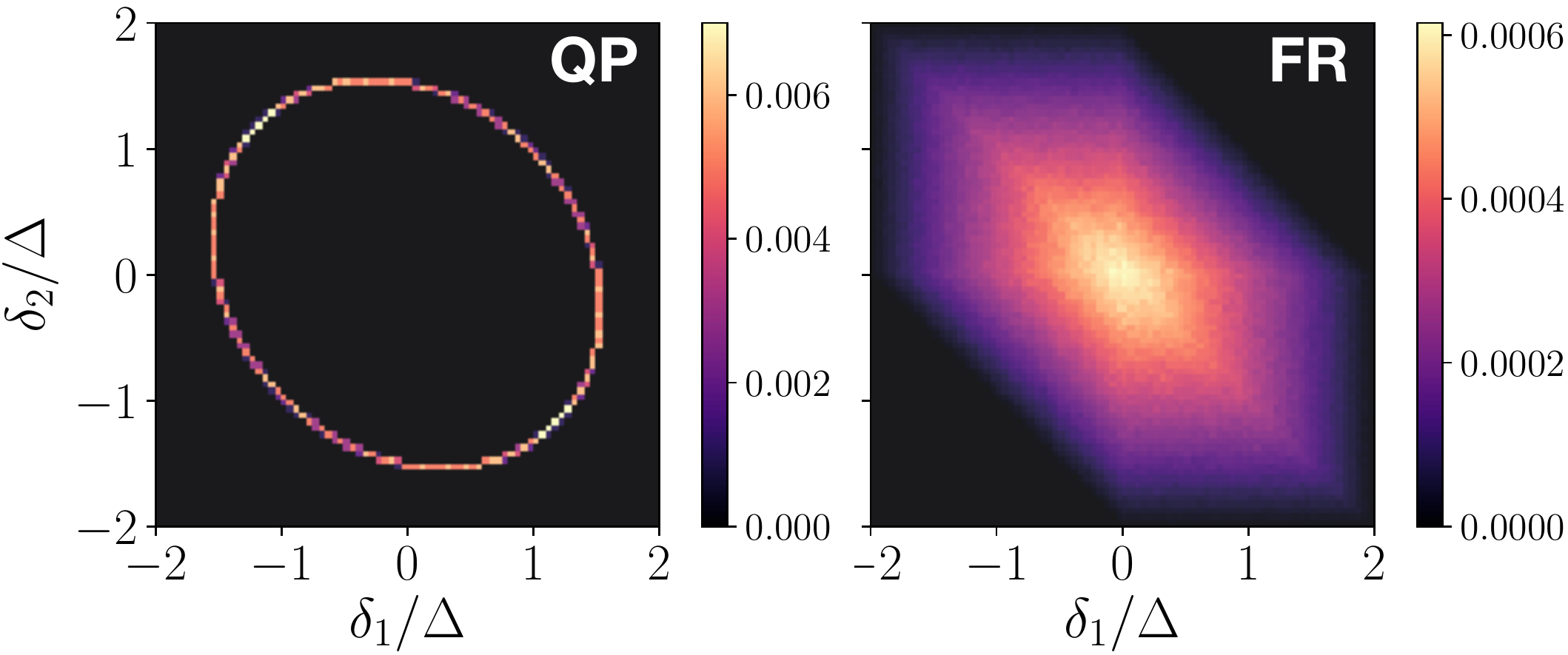}
    \caption{Numerically sampled probability distribution $P(\delta_1, \delta_2)$ for two energy differences $\delta_1$ and $\delta_2$ on contiguous pairs of sites. Left panel: QP potential; Right panel: FR potential.}
    \label{fig:Pd1d2}
\end{figure}

Such a form of correlation is indeed observed in the case of the quasiperiodic (QP) potential: figure~\ref{fig:Pd1d2}(a) shows the joint probability $P(\delta_1,\delta_2)$ for two adjacent energy differences, displaying a dip for $\delta_1 = \delta_2 =0$ -- an aspect which prevents two successive pairs of sites from being resonant simultaneously. In the case of interacting fermions, on the other hand, the above condition requires that if, \emph{e.g.} $J_z\pm \delta_1 \approx 0$, then $J_z\mp \delta_2$ is non-zero, or vice versa -- this prevents a state of the type $|101\rangle$ on the sites $(i,i+1,i+2)$ from being simultaneously (quasi-)resonant with $|110\rangle$ and $|011\rangle$, or, similarly, the state $|010\rangle$ from being quasi-resonant with $|100\rangle$ and $|001\rangle$. This is indeed guaranteed by the fact that $P(\delta,-\delta)$ is nearly vanishing for any finite $\delta$, except for $\delta\approx 1.25 \Delta$ -- but the latter situation does not lead to consecutive resonances when $\Delta > 1.25 J$, which is always the case in our study. 

On the other hand the uniform potential has no correlations between two consecutive energy differences, and the $P(\delta_1,\delta_2)$ distribution is the product of two triangular distributions for $\delta_1$ and $\delta_2$ -- shown in figure~\ref{fig:Pd1d2}(b). This implies that a fundamental assumption behind the 2LS model description is not guaranteed to be satisfied -- while it is more likely to have two adjacent pairs of sites with different energy offsets than with similar ones, one cannot exclude the existence of ``rare" regions with consecutive nearly resonant pairs. This requires to improve the 2LS model to a three-site (three-level) one, as detailed in section~\ref{sub:3LS}. 

\bigskip
\section{Loschmidt-echo dynamics for different disorder realizations}
\label{App:LEreal}
\begin{figure}[tb]
    \centering
    \includegraphics[width=\textwidth]{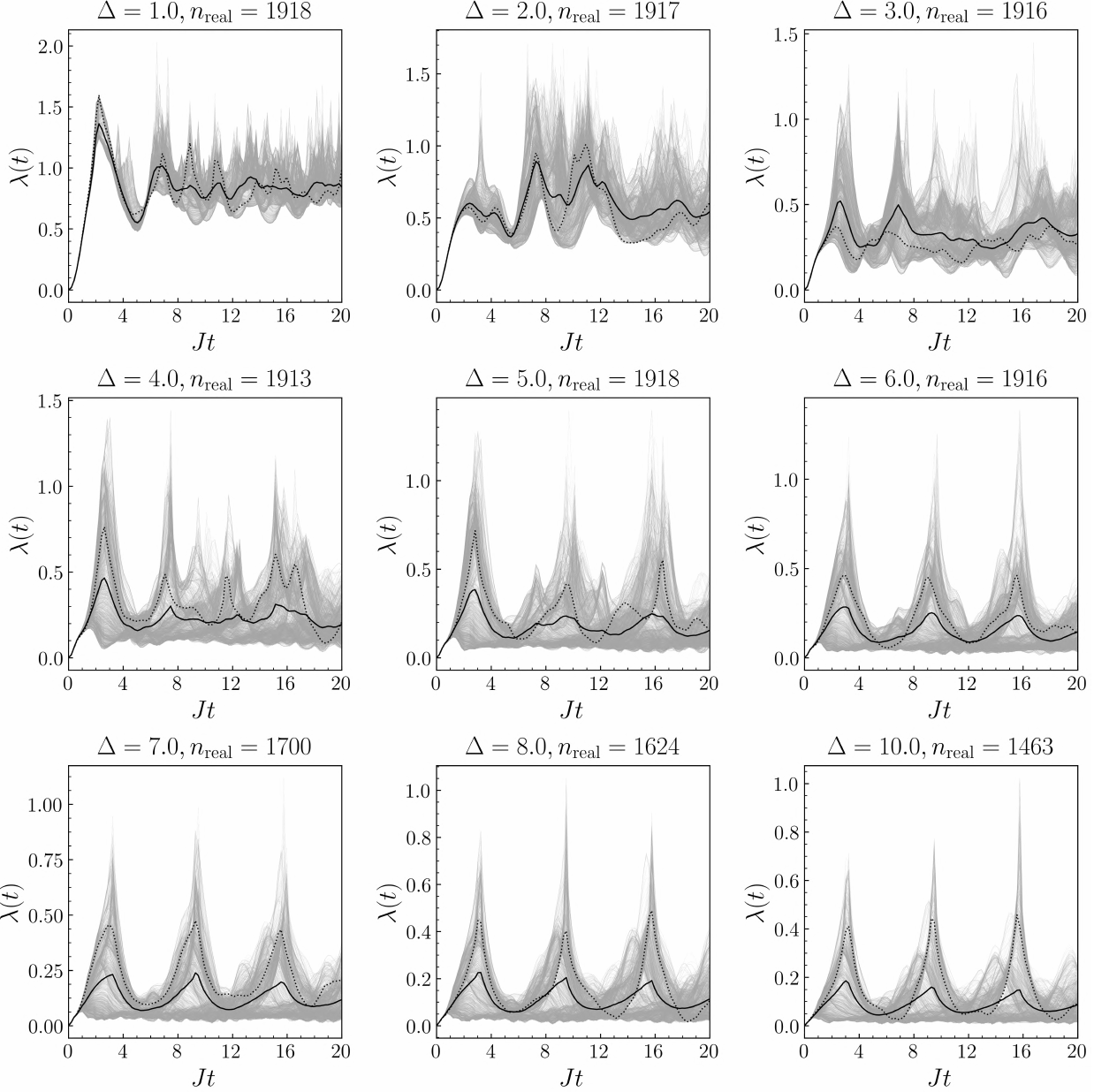}
    \caption{Averaged $\lambda(t)$ (black line) at different disorder strengths $\Delta=1,\dots, 10$ for a chain of $L=22$ sites in a QP, plotted along with all the realizations used for the averaging procedure (grey lines). The dotted line represents a typical individual realization exhibiting singular behavior.}
    \label{fig:QPreal}
\end{figure}

\begin{figure}[tb]
    \centering
    \includegraphics[width=\textwidth]{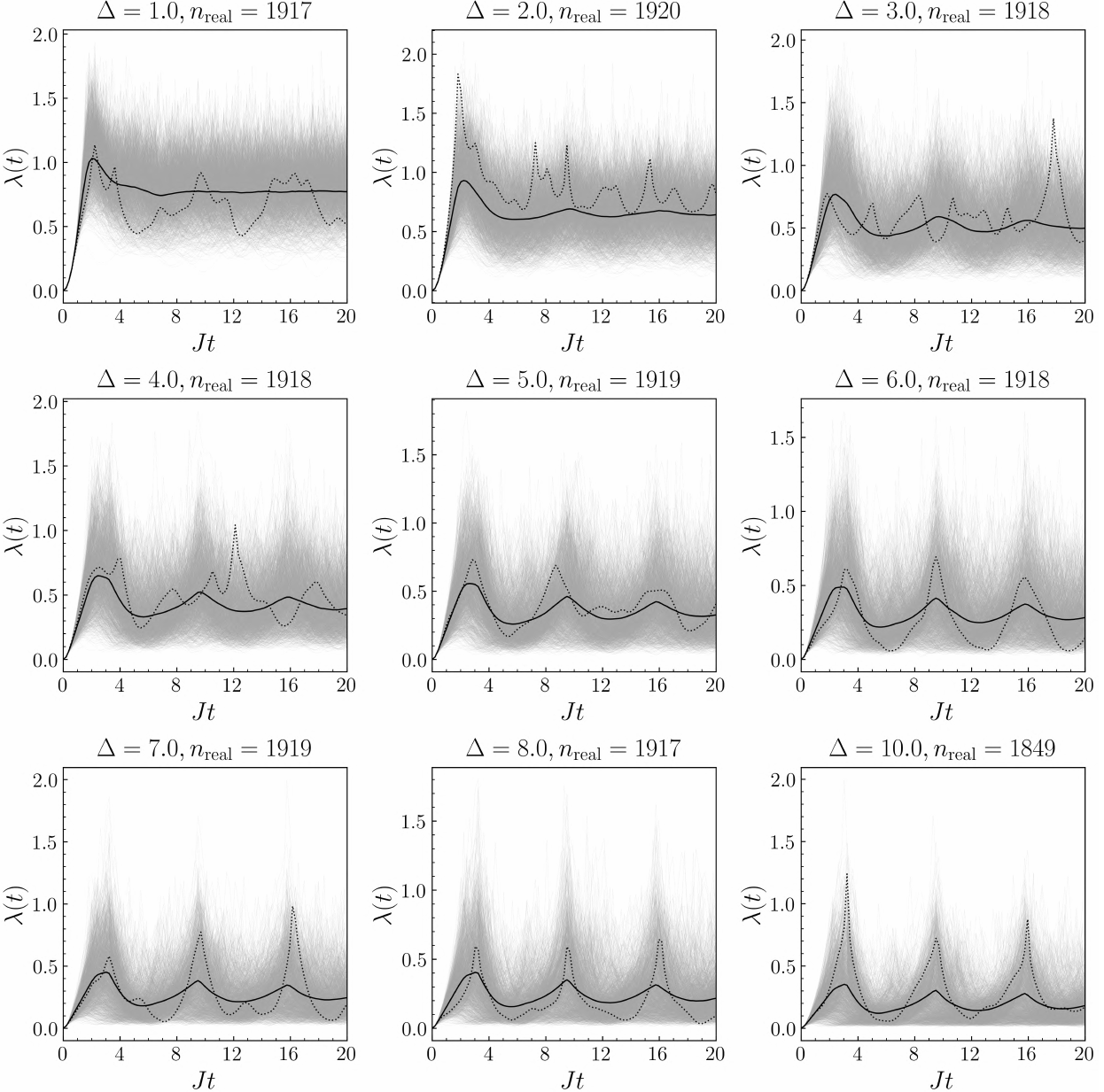}
    \caption{Averaged $\lambda(t)$ (black line) at different disorder strengths $\Delta=1,\dots, 10$ for a chain of $L=22$ sites in a FR potential, plotted along with all the realizations used for the averaging procedure (grey lines). The dotted line represents a typical individual realization exhibiting singular behavior.}
    \label{fig:FRreal}
\end{figure}

Figs.~\ref{fig:QPreal} and \ref{fig:FRreal} show the disorder average of the Loschmidt echo for a chain of $L=22$ sites, along with all the disorder realizations ($>10^3$) contributing the average, for various strengths ($\Delta/J = 1, 2, ..., 10$) of the QP and FR potential, respectively. We observe that sharp cusp singularities are exhibited by a signification portion of the disorder realizations, and that for sufficiently strong disorder these realizations are a finite fraction of the disorder statistics  (in the asymptotic limit), so that cusp singularities persist in the disorder-averaged results as well. These plots also suggest the fact that cusp singularities can be observed with a limited disorder statistics, under realistic experimental conditions. 

\end{document}